\documentstyle[12pt]{article}

\newcommand{\eq}{\begin{equation}}
\newcommand{\en}{\end{equation}}
\newcommand{\eqa}{\begin{eqnarray}}
\newcommand{\ena}{\end{eqnarray}}

\newcommand{\lbl}{\label}

\begin{document}

\hskip 11.5cm \vbox{\hbox{DFTT 36/93}\hbox{July 1993}}
\vskip 0.4cm
\centerline{\bf   THE C = 1 MATRIX MODEL FORMULATION}
\centerline{\bf   OF TWO DIMENSIONAL YANG-MILLS THEORIES}
\vskip 1.3cm
\centerline{ Stefano Panzeri}
\vskip .6cm
\centerline{\sl Istituto Nazionale di Fisica Nucleare, Sezione di Torino}
\centerline{\sl  Dipartimento di Fisica
Teorica dell'Universit\`a di Torino}
\centerline{\sl via P.Giuria 1, I-10125 Torino, Italy}
\vskip 2.5cm

\begin{abstract}
We find the exact matrix model description of two dimensional Yang-Mills
theories on a cylinder or on a torus and with an arbitrary compact
gauge group. This matrix model is the singlet sector of a $c =1$ matrix
model where the matrix field is in the fundamental representation of the
gauge group. We also prove that the basic constituents of the
theory are Sutherland fermions in the zero coupling limit, and this
leads to an interesting connection between two dimensional gauge
theories and one dimensional integrable systems. In particular we derive
for all the classical groups the exact
grand canonical partition function of
the free fermion system corresponding to a two dimensional gauge
theory on a torus.
\end{abstract}
\vskip 5cm
\hrule
\vskip1.2cm
\noindent

\hbox{\vbox{\hbox{$^{\diamond}${\it email address:}}\hbox{}}
 \vbox{\hbox{ Decnet=(31890::PANZERI)}
\hbox{ internet=PANZERI@TO.INFN.IT}}}
\vfill
\eject

\newpage

\section{Introduction}

Due to its property of exact solvability \cite{RuWitBlTh}, two
dimensional QCD (QCD2) may be a starting point for the analytical
investigation of the properties of the confined phase of four
dimensional QCD. Recently, some progress in this direction was done by
Gross and Taylor \cite{GT}. In fact they showed that QCD2 can be
interpreted, in the large $N$ limit, as a closed string theory, by
proving that all the coefficients of the expansion of the partition
function in powers of $1/N$ have a geometrical interpretation in terms
of maps of an orientable string worldsheet onto a two dimensional target
space.

Later, the nature of the string theory underlying QCD2 was further
clarified by finding the explicit matrix model description of QCD2, at
least on the cylinder and on the torus \cite{MiPo,Dou,CDMP}.
In \cite{MiPo,Dou} it was indeed shown that the hamiltonian for this
theory
in the large $N$ limit essentially coincides with the one found by
Das and Jevicki \cite{DJ} for the $c =1$ matrix model, the only
difference being that the corresponding matrix field is now a unitary
matrix field rather than an Hermitian one, while
in \cite{CDMP}  it
was found  that QCD2 on the torus and on the cylinder is, even
for finite $N$, exactly equivalent to a one dimensional
matrix model of type proposed in \cite{KM} by Kazakov and Migdal (KM),
where now the eigenvalues of the matter fields live on a circle rather
than on a line.

It is very interesting to understand  whether two dimensional Yang-Mills
theories with an arbitrary compact gauge group admit a similar closed
string theory formulation. In \cite{SOSP}  the large $N$
expansion of the partition function of YM2 with gauge group $SO(N)$ or $
Sp(2N)$ was studied, and  a string theory description similar to that of
\cite{GT} was obtained,
the main difference being that the worldsheet of the string
may be nonorientable.
In this letter we proceed in a different direction, by proving that YM2
with an arbitrary compact gauge group on a cylinder and  on a torus
are exactly the singlet sector of a $c =1$ matrix model whose matrix
field is in the fundamental representation of the gauge group and is
precisely the path-ordered integral of the gauge
field around the compactified
dimension. This result is obtained both in the continuum and in the
lattice formulation.

Further, it is discussed the free fermion content of the theory, by
showing that YM2 can be viewed as the zero coupling limit of the
Sutherland integrable systems \cite{Sut}. This free fermion
interpretation is the physical background for a Das-Jevicki matrix model
realization.
Finally, we remark that we obtain  for the partition function on the
torus (which corresponds to finite temperature Sutherland systems) a new
interesting expression in terms of Jacobi
theta functions, whose behaviour
under modular transformations is well known. The significance of the
modular inversion in the context of Sutherland systems is also
clarified.

All the results are presented for  simple compact Lie groups. The
generalization to an arbitrary compact group is easily done by taking
semidirect products.

The paper is organized as follows : in section 2 we derive the matrix
model action for YM2 and clarify its connection with the modular
inversion for the kernel on the cylinder; in section 3 we fully
exploit the fermion content of the theory and obtain
new expressions for the partition
functions on the torus; in section 4 we present our conclusions.

\section{The matrix model description of YM2}

It is well known by now that two dimensional Yang-Mills theories
 defined on a manifold
$\cal M $ of genus $p$ and with a metric $g_{\mu \nu} $ are exactly
solvable. The partition function is given by
\eqa
{\cal Z}_{\cal M}({\cal A}) & = & \int {\cal D} A_{\mu}
 e ^{- \frac{1}{4 \tilde{g}^{2}}
\int_{\cal M} d^{2} x \sqrt{g} {\rm Tr} F_{\mu \nu}
F^{\mu \nu}} \nonumber \\
 & = &
\sum_{R} d_{R}^{2-2p} e^{- \frac{1}{2} {\cal A} \tilde{g}^{2} C_{2}(R)} ,
\label{eq1}
\ena
where the sum is over all equivalence classes of irreducible
representations $R$ of the gauge group $G$,
$d_{R}$ is their dimension and $ C_{2}(R) $ is
the quadratic Casimir in the representation $R$.
Similarly, the heat kernel defined by a surface of genus $p$ and
$n$ boundaries is given by \cite{RuWitBlTh}
\eq
{\cal K}_{p,n} (g_1, \ldots , g_n; N, {\cal A}) = \sum_R
d_R^{2 - 2p - n} \chi_R(g_1) \cdots \chi_R(g_n)
e^{-\frac{1}{2} {\cal A} \tilde{g}^2 C_2(R)} ,
\label{heatker}
\en
where $g_i$ are the Wilson loops evaluated along the boundaries, and
$\chi_R$ denotes the Weyl character of the representation $R$.
For dimensional reasons, and because of the invariance of the action
under area preserving diffeomorphisms, eqs. (\ref{eq1}) and
 (\ref{heatker})
depend only on the variable  $\tilde{g}^2 {\cal A}$.
We will henceforth denote this variable by $t$.
Since in this paper we are dealing only with Kernels defined on $p=0$
surfaces, from now on we will denote simply $K_{0,n} $ by $K_n$.
The heat kernels in (\ref{heatker}) are expressed in terms of
exponentials of $g^2$, but
in order to formulate two dimensional YM theories on a general simple
Lie group
as a  matrix model,
we need to construct, as pointed out in \cite{CDMP},
a similar representation for the kernel defined by the cylinder
$({\cal K}_{0,2})$ in terms of exponentials of $1/g^2$.
This representation was first derived by Altschuler and Itzykson (AI)
\cite{AI} in the context of the study of multimatrix models by using
only the algebraic properties of characters. It may be very useful to
give a brief account of this proof, as follows.

First we have to fix some notations.
Let $G$ be a compact simple
Lie group (of rank $r$) and ${\cal G} $ its Lie algebra.
 Let ${\cal G}^C $ be the
complexification of $ {\cal G} $. Let $\langle , \rangle $ be an
invariant form on $ {\cal G}^C $ which is positive definite on
$i{\cal G}$.
Choose a Cartan subalgebra ${\cal H}^C$ of $ {\cal G}^C $, let
$ {\cal H} = {\cal H}^C  \cap {\cal G} $, and choose a set of positive
roots $\Sigma_{+} \subset i {\cal H} $. We identify
$ {\cal H}^C $ with its dual by means of $ \langle , \rangle $.
To each positive root $\alpha$ we associate the corresponding coroot
$ \hat{\alpha} = 2 \alpha / \langle \alpha , \alpha \rangle $.
The coroot lattice $ \hat{Q} $ is the lattice generated by the coroots.
Let $P$ be the weight lattice, which is the dual of $\hat{Q} $.

The analog of the Vandermonde determinant for the Hermitian matrices,
 for
$h \in  i{\cal H}^C $ is the polynomial
\eq
\Delta (h) = \prod_{\alpha \in \Sigma_{+}} \langle \alpha , h \rangle
\label{vand}
\en
It is the infinitesimal version of the Weyl's denominator,
\eq
\sigma ( h ) = e^{ i \langle \rho , h \rangle}
\prod_{\alpha \in \Sigma_{+}}
\left( 1 - e^{-i \langle \alpha , h \rangle } \right)
= \prod_{\alpha \in \Sigma_+} 2i \sin \left(
\frac{\langle \alpha , h \rangle}{2} \right)
\en
where $ \rho $ denotes
\eq
\rho = \frac{1}{2} \sum_{\alpha \in \Sigma_{+}} \alpha
\label{rho}
\en

The set of the highest weights of all the irreducible unitary
representations of $G$ (irreps in the following) is
\eq
P_{+} = P \cap \{ x \in i {\cal H} |
\langle x , \alpha \rangle \geq 0 , \alpha \in \Sigma_{+} \}
\label{irreps}
\en
Denote by $\chi_{\lambda} (g)$ and $d_{\lambda}$ the character  and the
dimension of the irrep corresponding to $\lambda \in P_{+}$.
Then the heat kernel on the disk is given by
\eq
{\cal K}_1 (g,t) =
\sum_{\lambda \in P_{+}} d_{\lambda} \chi_{\lambda}(g)
 e^{- C_2 (\lambda) t/2}
\en
where the quadratic Casimir $ C_2 (\lambda) $ has the following
expression :
\eq
C_2 (\lambda) = | \lambda + \rho |^2 - | \rho |^2
\en
At this point we have to compute the kernel on the cylinder :
\eq
{\cal K}_2 (g_1 , g_2^{-1} , t) =
\int_G dg {\cal K}_1(g_1 g g_2^{-1} g^{-1} , t) =
\sum_{\lambda \in P_+} \chi_{\lambda} (g_1) \chi_{\lambda}
( g_2^{-1} ) e^{- C_2 (\lambda) t/2 }
\label{Kcylsg}
\en
where  $dg$ denotes the normalized Haar measure on $G$ and
the latter expression is a consequence of the relation :
\eq
\int dg \chi_{\lambda} (x g y g^{-1}) = d_{\lambda}^{-1} \chi_{\lambda}
(x) \chi_{\lambda} (y)
\label{propcard}
\en
The definition of the Weyl character $\chi_\lambda$ is the following:
\eq
\chi_{\lambda} (g) = \frac{\nu_{\lambda + \rho} (\phi)}{\sigma(\phi)}
\label{Weylchar}
\en
where $ \phi_i ~ (i = 1, \ldots, r) $ are the invariant angles of
 $g ~ (g = e^{i\phi} ; ~ i\phi \in {\cal H} ) $ and
\eq
\nu_{\lambda} (\phi) =
\sum_{w \in W} \epsilon (w) e^{i\langle w(\lambda) , \phi \rangle }
\label{Weylnu}
\en
where $W$ is the Weyl group.
In order to express (\ref{Kcylsg}) in terms of theta functions, we have
to rewrite the sum over the irreps in the form of an
 unconstrained sum over the
whole weight lattice. This is easily done by substituting in
(\ref{Kcylsg}) the Weyl's formula (\ref{Weylchar}) and by noticing that
$ \nu_{\lambda}(\phi) = 0 $ if the stabilizer of $\lambda$ in $W$ is
non-trivial. The result is the following:
\eqa
\sum_{\lambda \in P_+} e^{-C_2 (\lambda) t/2} \nu_{\lambda + \rho}
(\phi) \nu_{\lambda + \rho} (\theta)
& = & \nonumber \\
e^{ | \rho |^2 t /2} \sum_{w \in W} \epsilon (w)
\sum_{\lambda \in P} e^{i \langle \lambda, \phi
- w(\theta) \rangle }
e^{- | \lambda |^2 t/2} & &
\label{AIlin}
\ena
where $\epsilon (w) = (-1)^{l (w)} $, $l(w) =$ length of
$w$ expressed as a product of Weyl's reflections and
we denote by $\phi$ [resp. $\theta$] the invariant angles of $g_1$
[resp. $ g_2$].
By means of the Poisson summation formula  we arrive finally at
(apart from some irrelevant multiplicative factor) :
\eqa
{\cal K}_2 (g_1,g_2^{-1}, t) & \equiv  &
{\cal K}_2(\phi,\theta,t)
\nonumber  \\   &  =  &
\frac{e^{| \rho |^2 t/2} (2\pi / t)^{ l/2}}{\sigma(\phi)
\sigma(- \theta)}
\sum_{w \in W} \epsilon(w) \sum_{\beta \in 2 \pi {\hat Q}}
e^{-| \phi - w(\theta) + \beta |^2 /2t}
\label{AIquad}
\ena
Recently, the formula (\ref{AIquad}) was derived, for the $SU(N)$
and $U(N)$ groups, in the context of the study of QCD2 \cite{CDMP}.
The main advantage of the derivation \cite{CDMP} is that it
allows to prove that QDC2 on a cylinder and on a torus is described
by a matrix model which is exactly a one dimensional KM model. The
formula (\ref{AIquad}) then arises by diagonalizing the matrix model and
by fixing the boundary conditions.
So, in order to find this $ c = 1 $ matrix model description also for
the general case of a
compact simple Lie group, it is needed to extend the
procedure worked out in \cite{CDMP} to the general case.

Recovering the matrix model action describing YM2 on a cylinder,
is a trivial extension of the
calculation done in \cite{CDMP} for QCD2 :  therefore we will outline
only the main ideas of the procedure \footnote{The coordinates on the
cylinder are denoted by $x,\tau$. Notice that we choose the time
direction (with coordinate $\tau$) as the compactified one. This notation
is opposite to that of \cite{MiPo,Dou}, but is more convenient for the
study of finite temperature lattice models considered below}.
By working in the first order
formalism, it is convenient to fix the gauge $\partial_0 A_0 = 0$; then
all the non static modes of the Fourier expansion (in the time
coordinate) of the fields can be integrated away and one obtains the
following expression for $\cal{K}_2$ :
\eqa
{\cal K}_2(g_{1},g_{2}^{-1},t) & = &
 \int {\cal D}B {\cal D}A e^{-\frac{\pi}{t}
{\rm Tr} \int_{0}^{2\pi} dx \left[ \partial B - i[A,B] \right]^{2}}
\times \nonumber \\
& \times &
\hat{\delta} \left(W(0),g_1\right) \hat{\delta}
\left(W(2 \pi),g_2^{-1}\right)
\psi(g_1) \psi(g_2^{-1}) .
\label{eq14}
\ena
where $B(x)$ and  $A(x)$ (matrix fields on the algebra)
denote the static modes of the $A_0(x,\tau)$ and $A_1(x,\tau)$
gauge fields respectively,
W(x) denotes the path-ordered integral of the
gauge field around the compactified
direction :
\eq
W(x) = {\cal P} e^{i \int_0^{2\pi} d\tau A_0(x,\tau)} ,
\label{eq5}
\en
and, in order to fix (resp. to $g_1$ and $g_2^{-1}~$)
 the values of  $W$ at the two
boundaries of the cylinder in a gauge invariant
manner, we have introduced
delta function acting on the space
of conjugation invariant functions (class functions),
denoted by $\hat{\delta}(g,h)$ and defined by
\eq
\hat{\delta} (g,h) = \int dU \delta (U g U^{-1} h) .
\lbl{delta}
\en
The factors $\psi (g_1)$ and $ \psi (g_2) $ are just normalization
factors; they depend only on the eigenvalues of $g_1$ and $g_2$ and they
will be chosen so that one obtains
the partition function on a torus
by identifying the two boundaries of the cylinder
and  by a group integration over the boundary conditions . It is
also important to remark  that, since the matrix field is
$W(x)$, the compactified time direction disappears from
this description and we have an exact dimensional reduction.

A less trivial step is the diagonalization of the model (\ref{eq14})
and the  recovering of the
AI formulas (\ref{AIlin},\ref{AIquad}). In order to do this,
one has to write, for a general compact simple group,  the
conjugation invariant
delta functions in (\ref{eq14}) in terms of periodic delta functions of
the invariant angles of the boundary conditions.
In order to obtain the latter expression, we note that by substituting
in (\ref{delta}) the character expansion for the delta functions and by
using the identity  (\ref{propcard}) one obtains
\eq
\int dg \delta (g e^{i\phi} g^{-1} e^{-i\theta} ) =
\sum_{\lambda \in P_+} \chi_{\lambda}(e^{i\phi})
\chi_{\lambda} ( e^{-i\theta})
\label{intstep}
\en
By writing explicitly
the Weyl's character formula
(\ref{Weylchar}) and by using the formula (\ref{AIlin})
evaluated in $t=0$, one finds finally the following expression :
\eqa
& & \int dg \delta (g e^{i\phi} g^{-1} e^{-i\theta} ) =
\nonumber \\
& = & \frac{1}{\sigma(\phi) \sigma(\theta)}
\sum_{w \in W} \epsilon (w) \sum_{\lambda \in P}
e^{i \langle \lambda , \phi - w(\theta) \rangle}
\nonumber \\
& = & \frac{1}{\sigma(\phi) \sigma(\theta)}
\sum_{w \in W} \epsilon (w) \sum_{\beta \in 2\pi \hat{Q}}
\prod_{i=1}^{N} \delta \left( (\phi - w(\theta) + \beta)_i
\right)
\label{deltaper}
\ena
where the latter relation is valid up to terms which are vanishing when
one performs the integral over $\phi,\theta$.
At this point,
by means of techniques which are by now standard in matrix models, the
matrix $B(x)$ can be diagonalized, the functional integral over its
eigenvalues performed and, with the help of the expression
(\ref{deltaper}) , one obtains the
Altschuler-Itzykson formula (\ref{AIquad}). In retrieving (\ref{AIquad})
the normalization factor $\psi$ has also been determined . It is given
by
$\psi(g) = \sigma(g) / \Delta(g) $.

Let's now clarify the string theory description arising from our
results : the action
in (\ref{eq14}) is a KM model in
one continuous dimension
(the spatial dimension of the cylinder) , where the role of KM gauge
field [resp. KM matter field] is played by $A(x)$ [resp. $B(x)$].
The main difference between (\ref{eq14}) and a KM model of the ordinary
type is that the boundary conditions are now depending on
$e^{2 \pi i B}$ rather than $B$.
Therefore the string theory interpretation of (\ref{eq14}) is
straightforward, although slightly different if we pass from the
cylinder to the torus.
Let us consider first YM2 on a torus : here the target space of the
matrix model (\ref{eq14}) is a circle and thus, exactly as the KM model
on a circle describes the singlet (vortex free)
sector of the $c = 1$ compactified
hermitian matrix model \cite{CAP,BouKaz}, YM2 on a torus turns out to be
the singlet sector of a matrix model on a circle whose matrix field is
defined on the group rather than on the algebra.
If YM2 lives on a cylinder, the target space of the corresponding matrix
model is a line. In this case the KM gauge field can be gauged away and
we have an ordinary $G$-matrix model on a line
(also in the singlet sector since the boundary conditions  are imposed
in a gauge invariant manner at the boundaries of this target space).

This is in complete agreement with the results obtained for $U(N)$ and
$SU(N)$ by Minahan and Polychronakos \cite{MiPo} : they found,
 by choosing the gauge $A_1(x,\tau) = 0$, that QCD2 on a
cylinder is equivalent to the singlet sector of a
$c = 1$ unitary matrix
model, where the unitary matrix field is just $ W = e^{2 \pi i B(x)}$.
However, their derivation is not valid for the case of the torus, where
it is not possible to fix the above mentioned  gauge.

Finally, we want to show that the matrix-model description
of YM2 simply  arises
also in the lattice formulation.
To see this, first consider a d-dimensional $G$-matrix model with
continuum action
\eq
S  =  - \beta ~ {\rm Tr}
 \int d^d x DU(x) DU^{-1}(x)
\label{contact}
\en
where $U$ is a matrix field in the fundamental representation of $G$.
The $D$ symbol denotes a covariant derivative operator and thus this
model is an extension of the KM one, where now the local degrees of
freedom are on the group rather than on the algebra. For $d =1$, as
discussed before, (\ref{contact}) is the matrix model formulation of YM2.
In order to recover the same result on the lattice, let's write the
discretization
of (\ref{contact})   :
\eqa
S  &  =  &
- \beta_0 {\rm Tr} ~
\sum_x \sum_{\mu = 1}^d \left(
U_x - V_{x;\mu} U_{x+\mu} V_{x;\mu}^{-1} \right)
\left(U_x^{-1} -V_{x;\mu} U_{x+\mu}^{-1} V_{x;\mu}^{-1} \right)
\nonumber  \\
  &  =  &   \beta_0 \sum_{x,\mu}
{\rm Re} ~ {\rm Tr}
\left( U_{x} V_{x;\mu} U_{x+\mu} V_{x;\mu}^{-1} -2 \right)
\label{latact}
\ena
where $\beta_0 = \beta/a^{3-d}$ and $a$ is the lattice spacing.
The  KM matter fields on the sites $x$
of a regular hypercubic lattice are now $U_x$, while
the angular variables on the links are $V_{x+\mu}$. We shall call this
lattice field theory the KM G-matrix model.

It is interesting to note that, by thinking $U_x$ as a link matrix along
an extra compactified timelike direction, (\ref{latact})
is just the Wilson action for the
$d+1$ dimensional YM theory at finite temperature, where we are keeping
only the timelike plaquettes. In the timelike
direction we have actually only one site, but, at least in the large
$\beta$ limit, this is essentially irrelevant \cite{CAPht}.
As further discussed in \cite{CAPht},
this model should capture most of the
features of the infinite temperature limit of Yang-Mills theories.
In this context, for $d > 1$, $\beta$ plays the role of the coupling for
timelike plaquettes, and is proportional to the temperature.

At this point the diagonalization procedure is very similar to that of
the KM model, and one remains only with the invariant angles $\phi_x$
of the $U(x)$ matrices on the sites.
The analog of the Itzykson-Zuber formula
for the integration over angular link variables is
\eqa
I(\phi_x , \phi_{x+\mu}) &  =  &
\int_G  D V_{x;\mu} ~ \exp \left[  \beta_0  {\rm Re} {\rm Tr}
U_x V_{x;\mu} U_{x+\mu}^{-1} V_{x;\mu}^{-1} \right]
\nonumber  \\
&  =   &    \sum_R \lambda_R(\beta_0 )  \chi_R(e^{i \phi_x})
\chi_R (e^{-i\phi_{x+\mu}})
\label{intlink}
\ena
where we used the character expansion of Wilson action. The coefficients
of this expansion are :
\eq
\lambda_R(\beta_0 ) \equiv  d_R^{-1} \int_G  dU
\chi_R(U) e^{\beta_0  {\rm Tr} (U + U^{-1})}
\label{lamc}
\en
and have the following asymptotic behaviour as
$\beta_0 \rightarrow \infty$ :
\eq
\lambda_R(\beta_0) ~  \sim ~  1 - C_2(R)/2\beta_0  + O(1/ \beta_0^2)
\label{aslam}
\en
Here we are interested only to the $d=1$ model,
since we want to show that it is equivalent to YM2 on a cylinder.
 In the $d =1$ case, no
approximation is needed in keeping only the timelike plaquettes in the
Wilson action for YM theories (there is no spacelike plaquette
contribution)
 and the continuous limit $(\beta_0 \rightarrow \infty)$
 is well defined. So, let's consider in detail the discretization of YM2
with Wilson action on a cylinder, and denote with $N_0$ the number of
sites in the compactified time
direction and with $N_1$ the number of sites in the
space direction. The boundary conditions are assigned by fixing
in a conjugation invariant manner the value of the Polyakov loop which
winds around the compactified direction. We can actually, by using
the orthogonality properties of characters,
reduce ourself to the $N_0 = 1$ case (This leads, in the continuous
limit, only to a rescaling of the coupling constant $\beta \rightarrow
\beta/N_0 $). At this point we remain with a KM G-matrix model on a one
dimensional lattice with $N_1$ sites, which is thus equivalent in the
continuous limit to YM2 on the cylinder.
In the $\beta_0 \rightarrow \infty$ limit, it is also easy
to obtain from (\ref{intlink},\ref{lamc},\ref{aslam}) the usual
continuum
expression for the kernel on the cylinder.

Notice that the G-matrices on
the sites  are  the lattice counterparts of
the path-ordered integrals of the  gauge field
around the time direction. The extension to
the case of the torus is straightforward, the only
difference being that the
KM G-matrix model is defined on a circle.

\section{The free fermion content of YM2}

In this section we show that, as expected from the matrix model
interpretation \cite{Poly},
YM2 can be viewed as a quantum theory of free fermions
living on a circle. More precisely, it is the zero coupling limit of the
Sutherland integrable systems \footnote{This correspondence was first
noticed, in a different context, in \cite{GN}.}
 \cite{Sut}.
(For a comprehensive review see \cite{OlPe})

One can start by noticing, in analogy with the well known analysis for
Hermitian $c =1$ matrix model \cite{BIPZ}, that the quantization of
(\ref{eq14}) is equivalent to the singlet sector of G-matrix quantum
mechanics. Thus the hamiltonian written in terms of the invariant angles
on G is :
\eq
H = - \Delta_G \equiv
-  \sigma^{-1}(\phi)  \left[  \sum_i
\frac{\partial^2}{\partial \phi_{i}^{2}} + | \rho |^2
\right] \sigma(\phi)
\label{Ham}
\en
where $\Delta_G$ is the ``radial'' part
of the Laplace operator on the
group manifold \cite{Dow} and $\rho$ is defined in (\ref{rho}).
This is a consequence of the fact that the
kernels (\ref{heatker}) of YM2, for example the kernel on the disk
${\cal K}_1$, are solutions of the heat equation :
\eq
\left( \frac{\partial}{\partial t} + \frac{1}{2} \Delta_G
\right) {\cal K}_1 (\phi,t)  =  0
\label{heateq}
\en
After redefining the kernel by ${\cal K}_1 \rightarrow \sigma(\phi)
{\cal K}_1(\phi,t) \equiv \hat{\cal K}_1(\phi,t)$,
the eq. (\ref{heateq}) becomes the
(euclidean) free Schr\"{o}dinger equation for $r$ fermions on a circle.
It is important to notice that the boundary conditions are determined by
this redefinition, and are just the same boundary conditions for the
fermions in the zero coupling limit of the Sutherland integrable model
related to the Lie algebra ${\cal G}$ \cite{OlPe}.

As a consequence one can rewrite all the quantities of YM2 in the
language of the Sutherland model. In fact, the invariant angles $\phi_i$
on the gauge group G are the coordinates of the fermions; the
eigenvalues of the hamiltonian are $C_2(\lambda)$ (the spectrum is
discrete because of the compactness of the configuration space); the
eigenfunctions corresponding to $C_2(\lambda)$ are the class functions
$\psi_{\lambda}(\phi) = \psi_0 (\phi) \chi_{\lambda}(\phi)$,
where $\psi_0(\phi) \equiv \sigma(\phi)$ is the ground state
eigenfunction.
The transition amplitude  between the configuration
$\phi_i =0$ at $t=0$ and the configuration $\phi_i$ at time $t$ is
proportional to the kernel on the disk, and it is given by
$\hat{\cal K}_1$.
Similarly the propagator  from the configuration
$\phi_i$ at $t=0$ to the configuration $\theta_i$ at time $t$ is
essentially the kernel on the cylinder :
\eq
\hat{\cal K}_2 (\phi, \theta, t) =
\sigma(\phi) \sigma(-\theta) {\cal K}_2(\phi,\theta,t)
\label{pks}
\en
It is now clear the meaning, in the Sutherland  system language,
of the modular inversion leading to the expression (\ref{AIquad}) for
${\cal K}_2$ : the usual character expansion (\ref{AIlin})
corresponds to
the momentum representation for the Sutherland propagator (in fact the
elements of the weight lattice labels the momentum eigenvalues), while
the inverted expression (\ref{AIquad})  is a gaussian in the invariant
angles and thus corresponds to the coordinate representation for the
fermion propagator (the integers numbers which label the coroot lattice
in (\ref{AIquad}) have to be interpreted as winding numbers).
In \cite{OlPe} it is shown that the
Sutherland model can be related to the
Laplace operator $\Delta_G$ also for a set of nonzero values of the
coupling constant. The formula (\ref{AIquad}) provides, similarly as the
noninteracting case, a modular inversion for the one and two point
functions also in this cases.

Even more interesting is the interpretation of the YM2 partition
function on a $p=1$ surface, obtained by simply sewing together the two
ends of the cylinder, according to \footnote{Notice that in
(\ref{sewing})
we have ignored, for simplicity, the overall factor $\exp [ |
\rho |^2 t/2 ]$ and this corresponds  just
to a shift of the zero point
energy \cite{Het}.}:
\eq
{\cal Z}_{p=1}(t) = \int dg {\cal K}_{2}(g,g^{-1},t) =
\int_0 ^{2\pi} \prod_{i=1}^{r} d \phi_i
\sum_{w \in W} \epsilon (w)
\sum_{\lambda \in P}
e^{i \langle \lambda , \phi - w(\phi) \rangle }
e^{- | \lambda | ^2 t/2}
\label{sewing}
\en
where with $\phi_i$ we denote the  invariant angles of $g$.
Since we are now fixing periodic boundary conditions in the
euclidean time direction, the partition function on the torus
(\ref{sewing})  is just  the finite temperature version of the
corresponding Sutherland model, the inverse temperature  being
proportional to $t = g^2 \cal{A}$.
Here we want to show that, as suggested in \cite{CDMP} for the $SU(N)$
group, it is possible to find for the partition function on the torus an
alternative expression, different from the usual character expansion.
The interest of this result is twofold : on one hand we
have an expression for ${\cal Z}_{p=1}$ in terms of theta functions
$\theta_2$ and $\theta_3$,
whose behaviour under modular transformations is well known; on the
other hand this allows us to write in a new way the energy levels of the
zero coupling limit (at finite temperature) of the Sutherland systems
related to a given Lie algebra.
Since this expressions are very simple in the  grand canonical
formalism, we restrict ourself to
the classical simple groups \footnote{For the exceptional groups, it is
obviously not possible to use the grand-canonical formalism; however one
can still write the canonical partition function as a sum of a finite
number of theta functions.}.

In the case of the classical simple groups, the Weyl group
$W$ has the permutation group of the invariant angles as a subgroup.
This suggests us to calculate the above integral by decomposing,
as done in \cite{CAP} for the one-dimensional KM model,
each permutation belonging to $W$ into its cycles.
At this point we have to consider separately the four series  of
classical simple groups : $Sp(2N)$, $SO(2N+1)$,
$SO(2N)$, and $SU(N)$ (corresponding to the four series of
Lie algebras
$C_N$, $B_N$, $D_N$ and $A_N$).

\vskip .6cm
{\bf 3.1 The partition function for Sp(2N)}
\vskip 0.3cm

The $Sp(2N)$ group has rank $N$ and the Weyl group is the group of
the permutations and sign changes of $N$ elements.
By writing explicitly the structure  of the weight lattice
$P$ and the structure of $W$ (as a
sum over permutations $S$ and sign changes
$\epsilon_i$),
the partition function (\ref{sewing}) has the form :
\eqa
{\cal Z}_{p=1}(N,t) & = &
\int_0 ^{2\pi} \prod_{i=1}^r d\phi_i
\sum_{\epsilon_i = \pm 1}
\sum_S
(-1)^S \prod_i \epsilon_i
\sum_{ \{ l_i \} = - \infty}^{+ \infty} \times \nonumber  \\
& \times &
\exp
\left\{ i \sum_k l_k ( \phi_k - \phi_{S(k)} \epsilon_{S(k)})
 - \sum_k l_k^2 t/2 \right\}
\label{plsp}
\ena
The Weyl group is larger than the permutation group;
if one first performs the sum over sign changes in (\ref{plsp}),
then as one decomposes each permutation $S$ into product of cycles, each
integral at the r.h.s. of (\ref{plsp}) decomposes into the product of
integrals corresponding to the cycles in the decomposition of $S$. In
this way (see \cite{CAP} for the details of cycle decomposition) one can
show that
\eq
{\cal Z}_{p =1}(N,t) =
\int_0^{2\pi}
\frac{d\theta}{2\pi}
\exp \left\{  -iN\theta - \sum_{j =1}^{\infty} \frac{(-1)^j}{j}
F_j e^{i\theta j} \right\}
\label{cyclestheta}
\en
With  $F_j$ we denote a contribution from a cycle of a length $j$;
it is given by :
\eqa
F_j & = &
\sum_{\epsilon_i = \pm 1}  \prod_{i=1}^j \epsilon_i
\int_0^{2\pi} d\phi_i
\sum_{\{l_i\} = -\infty}^{\infty}
e^{i\sum_k l_k (\phi_k - \phi_{k+1} \epsilon_{k+1} )}
e^{-\sum_k l_k^2 t/2}
\nonumber  \\
& = &
(2\pi)^j ~ 2^{j-1} ~ \left[ \theta_3 (0 , \tau j )
- 1 \right]
\label{Fjsp}
\ena
where from now on $\tau = \frac{it}{2\pi}$.
This result can be expressed in a rather elegant and interesting form
if one consider the grand-canonical partition function for the $C_N$
series :
\eq
{\cal Z}_{C}(q,t) = \sum_N {\cal Z}_{p=1}(N,t) q^N
\en
In this case one obtains :
\eq
{\cal Z}_{C}(q,t) = \exp \left\{ - \sum_{j=1}^{\infty}
\frac{(- 4 \pi q)^j}{2j} \left[ \theta_3 (0 , \tau j)
  -1 \right] \right\}
\en
The sum over $j$ in the exponents can be performed if one replaces
the theta functions with their expressions as infinite sums, finally
leading to the result
\eq
{\cal Z}_{C}(q,t) =   \prod_{n=1}^{\infty}
\left( 1 + 4\pi q e^{-\frac{t}{2} n^2} \right)
\label{gasfermionssp}
\en
In this infinite product we easily recognize the grand-canonical
partition function of a gas of free fermions at finite temperature and
in a compactified one-dimensional space.
\vskip .6cm
{\bf  3.2 The partition function for SO(2N+1)}
\vskip .3cm

The $SO(2N+1)$ group has rank $N$, the
 Weyl group is isomorphic to that
of $SP(2N)$ and the sum over the weight lattice
is given by two independent sums: the first sum corresponds to the
vectorial representations (labeled by integers numbers), while the
second sum corresponds to the spinorial ones (labeled by half-integers
numbers).
Therefore  the partition function
is conveniently written as the sum of two integrals:
\eq
{\cal Z}_{p=1}(N,t) = {\cal Z}_{p=1}^{(1)}(N,t)
+ {\cal Z}_{p=1}^{(2)}(N,t)
\label{Zsepsod}
\en
where
\eqa
{\cal Z}_{p=1}^{(1)}(N,t) & = &
\int_0 ^{2\pi} \prod_{i=1}^r d\phi_i
\sum_{\epsilon_i = \pm 1}
\sum_S
(-1)^S \prod_i \epsilon_i
\sum_{ \{ l_i \} = - \infty}^{+ \infty}
\times \nonumber \\
& \times &
\exp \left\{
i \sum_k l_k \left( \phi_k - \phi_{S(k)} \epsilon_{S(k)} \right)
 - \sum_k l_k^2 t/2 \right\}
\nonumber \\
{\cal Z}_{p=1}^{(2)}(N,t) & = &
\int_0 ^{2\pi} \prod_{i=1}^r d\phi_i
\sum_{\epsilon_i = \pm 1} \sum_S
(-1)^S \prod_i \epsilon_i
\sum_{ \{ m_i \} = - \infty}^{+ \infty}
\times \nonumber  \\
&  \times   &
\exp \left\{ i \sum_k \left( m_k + \frac{1}{2} \right)
 \left( \phi_k - \phi_{S(k)} \epsilon_{S(k)} \right)
 - \sum_k \left( m_k + \frac{1}{2} \right)^2 t/2 \right\}
\label{Zlinsod}
\ena
are the contribution to the partition function on the torus arising
respectively from vectorial and spinorial representations.
Each of the two terms in (\ref{Zsepsod}) can be computed as before;
the contributions of a cycle of length $j$ to the cyclic decomposition
of $S$ are then :
\eqa
F_j^{(1)} & = &
(2\pi)^j ~ 2^{j-1} ~ \left[ \theta_3(0 , \tau j ) -1 \right]
\nonumber  \\
F_j^{(2)} & = &
(2\pi)^j ~ 2^{j -1} ~ \theta_2 (0 , \tau j)
\label{Fjsod}
\ena
Thus the generating function $ {\cal Z}_{B} $
 for the $SO(2N+1)$ partition functions on a
torus has the form :
\eqa
{\cal Z}_{B}(q,t) & = &
\exp \left\{ - \sum_{j=1}^{\infty}
\frac{(-4\pi q)^j}{2j} \left[ \theta_3(0 , \tau j ) - 1
\right] \right\}
\nonumber  \\
&  +  & \exp \left\{ -
\sum_{j=1}^{\infty} \frac{(-4\pi q)^j }{2j}
\theta_2 (0 ,\tau j) \right\}
\label{zgctsod}
\ena
or equivalently
\eq
{\cal Z}_{B}(q,t) =
\prod_{n=1}^{\infty} \left( 1 + 4\pi q e^{-n^2 t/2} \right) +
\prod_{n=0}^{\infty} \left( 1 + 4\pi q e^{-(n+\frac{1}{2})^2 t/2} \right)
\label{gasfermionssod}
\en

\vskip .6cm
{\bf 3.3  The partition function for SO(2N)}
\vskip .3cm

By looking at the $SO(2N)$ group, we note that the weight lattice has
the same structure as that of $SO(2N+1)$, but the Weyl group is
different : it is the group of the permutations and sign changes of $N$
objects, but with the constraint that the total number of minus signs
must be even.
Therefore the partition function has a form similar to (\ref{Zsepsod},
\ref{Zlinsod}), the only difference being that the sum over the sign
changes is not the sum over the $N$ independent $\epsilon_i$,
but they do satisfy the relation $\prod_i \epsilon_i = 1$.

By writing , as in eq. (\ref{Zsepsod}), ${\cal Z}_{p=1}(N,t)$ as
a sum of two pieces and by decomposing each term into cycles one
obtains:
\eqa
{\cal Z}_{p=1}^{(i)}(N,t) &  = &
\sum_{m=0}^{\infty} \int_0^{2\pi} \frac{d\varphi}{2\pi}
e^{-2im \varphi} \int_0^{2\pi} \frac{d\theta}{2\pi}
e^{-iN \theta} \times \nonumber \\
& \times & \exp \left\{ - \sum_{j=1}^{\infty}
\frac{(-1)^j}{j} F_j^{(i)} e^{i \theta j} \right\}
 \label{cyclsop}
\ena
where $i = 1,2$ and
\eqa
F_j^{(1)} & = &
(2\pi)^j ~ 2^{j-1} ~ \left[ \theta_3 (0, \tau j) - e^{i\varphi}  \right]
\nonumber  \\
F_j^{(2)} & =   & (2\pi)^j ~ 2^{j-1} ~
\theta_2 (0, \tau j ) \label{Fjsop}
\ena
The sum over the positive integers $m$  and the integral over the
Lagrange multiplier $\varphi$ has been inserted in
(\ref{cyclsop},\ref{Fjsop}) in order to constraint the sign changes
variables to satisfy the above mentioned relation.
Now we can write the grand-canonical partition function :
\eqa
{\cal Z}_{D}(q,t) & = &
\left( 1 + 2\pi q \right)
\prod_{n=1}^{\infty} \left( 1 + 4 \pi q e^{- n^2 t/2} \right)
\nonumber   \\
& + & \prod_{n=0}^{\infty} \left( 1 + 4 \pi q
e^{- (n + \frac{1}{2} )^2 t/2} \right)
\label{gasfermionssop}
\ena
It is interesting to remark that, for the orthogonal groups, due to the
contribution of spinorial representations, the winding numbers (i. e.
the integers numbers labeling the coroot lattice) in the Sutherland
propagator (\ref{AIquad}) are only even, as usual when we are dealing
with fermionic representations.

\vskip .6cm
{\bf 3.4  The partition function for SU(N)}
\vskip .3cm

Let us conclude this section
by remembering that for $SU(N)$ this procedure was
worked out in ref. \cite{CDMP}. For the sake of completeness,
we report here the expression of ${\cal Z}_{A}$ that one finds
in this case :
\eqa
{\cal Z}_{A}(q,t) &  = &
 \left( \frac{t}{4\pi} \right)^{1/2} \int_{0}^{2
\pi}
d\beta \prod_{n=-\infty}^{+\infty} \left( 1 + q
e^{- \frac{t}{2} \left( n - \frac{\beta}{2\pi}
\right)^2 } \right) .
\label{gasfermionsSUN}
\ena
The $SU(N)$ group has actually rank $N-1$, although its root lattice is
most conveniently embedded in a $N$ dimensional space. This fact, for
the system of fermions, means that the wave function of the center of
mass of the $N$ fermions is completely localized, and therefore the
corresponding momentum undetermined. This is the reason of the
integration over $\beta$ in (\ref{gasfermionsSUN}).

Finally we emphasize that the grand canonical partition functions has
been defined keeping $t$ independent of $N$, so all
these grand canonical
expressions cannot be used as such to calculate the large $N$ limit of
${\cal Z}_{p=1}(N,t)$

\section{Conclusions}

We have shown in this paper that YM2 on a cylinder or on a torus,
with gauge group an arbitrary compact  Lie group $G$ is exactly
a one dimensional matrix model of the type proposed by
Kazakov and Migdal, where the KM matter field is now in the fundamental
representation of the gauge group. The string interpretation of this
model requires expanding the group matrices as exponentials of the
matrices on the algebra; therefore in our formulation it is quite direct
to retrieve the result found in \cite{GT,SOSP} that
the worldsheet is orientable for the $SU(N)$ gauge group, while it may
be both orientable and nonorientable for YM2 with gauge group $SO(N)$
or $Sp(2N)$.
However, we are not able to extend these results
to YM2 on a higher genus surface.

We also prove that the fundamental constituents of the theory are free
fermions excitations, most conveniently described as the zero coupling
limit of the Sutherland systems and  it is very intriguing, for future
developments, to note that the large $N$ limit
of these integrable systems, as shown in
\cite{KaYa} for the Sutherland systems related to $SU(N)$,
is described by a two dimensional conformal field theory.

\vskip 1cm
I am very indebted with M. Caselle and A. D'Adda for daily
discussions,
for many suggestions and for a careful reading of the manuscript. I
would like also to thank M. Bill\`{o}, F. Gliozzi
and L. Magnea for useful discussions.

 \end{document}